\documentclass[aps,prd,preprint,showpacs,superscriptaddress]{revtex4}

\begin{document}
\title{Can the Majoron be gauged away?}
\author{Luis N. Epele}
\affiliation{Laboratorio de F\'\i sica Te\'orica, Departamento de
F\'\i sica, Universidad Nacional de La Plata, C.C. 67-1900, La Plata
Argentina.}
\author{Huner Fanchiotti}
\affiliation{Laboratorio de F\'\i sica Te\'orica, Departamento de
F\'\i sica, Universidad Nacional de La Plata, C.C. 67-1900, La Plata
Argentina.}
\author{Carlos Garc\'\i a Canal}
\affiliation{Laboratorio de F\'\i sica Te\'orica, Departamento de
F\'\i sica, Universidad Nacional de La Plata, C.C. 67-1900, La Plata
Argentina.}
\author{William A. Ponce}
\affiliation{Instituto de F\'\i sica, Universidad de Antioquia,
A.A. 1226, Medell\'\i n, Colombia.}

\begin{abstract}
{In a recent paper [Phys. Rev. D\textbf{73}, 075005 (2006)], the
authors presented the lepton number violation mechanisms in the
econimic version of the 3-3-1 model, without explicitly pointing where the
pseudo Goldstone Majoron lies. In this comment we clarify this
point and show an extended version of the model where the
Goldstone Majoron becomes a gauge away would be Majoron.}
\end{abstract}



\maketitle

Masses for neutrinos require physics beyond the standard model
(SM) connected either to the existence of right handed neutrinos
and/or to the breaking of the Barion minus Lepton (B$-$L)
symmetry. Besides, Majorana masses for neutrinos require the
violation of the lepton number; violation that can be spontaneous
implying the existence of the so called Majoron~\cite{gelron}, or either
explicit implying the existence of a pseudo Goldstone Majoron.

Interesting extensions of the SM are based on the local gauge
group $SU(3)_c\otimes SU(3)_L\otimes U(1)_X$ (3-3-1 for short) in
which the weak sector of the SM is extended to $SU(3)_L\otimes
U(1)_X$. Several models for this gauge structure have been
constructed so far, being the most popular the original
minimal model~\cite{pf} and the so called model with right handed
neutrinos~\cite{long}. An amusing variant of this last model is
the so called ``economic 3-3-1 model"~\cite{ponce, lon, malong} in which
the spontaneous symmetry breaking and the generation of masses for
all the particles in the model is done by a minimal set of only
two Higgs scalar triplets.

One, among the many outstanding features of 3-3-1 models, is the
appropriate explanation of neutrino masses and oscillations. The
particular analysis done for the minimal version of the model, and
for the model with right handed neutrinos, carried at the expense
of violating the lepton number L by two units, implies in some
cases the existence of a ruled out Majoron~\cite{gelron}. But the
economic version of the model presents special features as
discussed in Refs.~\cite{lon, malong} and in what follows.

The fermion content of the economic model is given by the
following 3-3-1 anomaly free structure~\cite{ponce} :
$\psi_{lL}^T= (\nu_l^0,l^-,\nu_l^{0c})_L\sim (1,3,-1/3),\;
l_L^+\sim(1,1,1),\; Q_{iL}^T=(d_i,u_i,D_i)_L\sim (3,3^*,0),\;
Q_{3L}^T=(u_3,d_3,U)_L\sim (3,3,1/3)$, where $l=e,\mu,\tau$ is a
family lepton index, $\nu_{lL}^{0c}$ stands for the right-handed
neutrino field and $i=1,2$ for the first two quark families. The
right handed quark fields are $u_{aL}^c\sim (3^*,1,-2/3),\;
d_{aL}^c\sim (3^*,1,1/3),\; D_{iL}^c\sim(3^*,1,1/3)$ and
$U_L^c\sim (3^*,1,-2/3)$, where $a=1,2,3$ is the quark family
index and there are two exotic quarks with electric charge $-1/3\;
(D_i)$ and another with electric charge 2/3 $(U)$.

The Gauge Boson sector of the model includes, besides $W^\pm$ and
$Z^0$ as in the SM, five more particles: $K^\pm,\; K^0,\;
\overline{K^0}$ (which carry U and V isospin respectively) and an
extra Boson $Z^{\prime 0}$ related to a second diagonal weak
neutral current. The scalar sector consists of only two scalars
triplets (instead of three as in the original model with right
handed neutrinos~\cite{long}) which are:
$\chi^T=(\chi_1^0,\chi_2^-,\chi_ 3^0)\sim (1,3,-1/3)$ and
$\rho^T=(\rho_1^+,\rho_2^0,\rho_3^+)\sim (1,3,2/3)$, with the
following Vacuum Expectation Values (VEV)
$\langle\chi\rangle^T=(v_1,0,V)$ and
$\langle\rho\rangle^T=(0,v,0)$, where~\cite{ponce} $v_1<v\sim 10^2$ GeV$<V\sim$ TeV.

The lepton number L is not a good quantum number in the context of
this model because, $\nu^{0c}_{lL}$, the third component in the
triplet $(1,3,-1/3)$, is identified with the antiparticle of
$\nu^0_{lL}$, which in turn implies that L does not commute with
$SU(3)_L\otimes U(1)_X$. The lepton number assignments for this
model are~\cite{lon}: L$(l^-_L,\;
\nu^0_{lL})=-$L$(l^+_L,\;\nu^{0c}_{lL})=1$, L$(u_{aL},\;
u_{aL}^c,\; d_{aL},\; d_{aL}^c,\; W^\pm ,\; Z^0,\; Z^{\prime
0})=0$; L$(K^{+},\; K^0,\; U_L,\; D_{iL}^c)=-2=-$L$(K^-,\;
\overline{K}^0,\; D_{iL},\; U_L^c)$. Then, the Yukawa coupling
constants imply L$(\chi_2^-,\;\chi^0_1)=2$, L$(\rho_3^+,)=-2$, and
L$(\rho_1^+,\; \rho_2^0,\;\chi_3^0)=0$. Notice that $D_i^{-1/3}$
and $U^{2/3}$ are bileptoquarks and $K^+,\; K^-,\; K^0$ and
$\overline{K}^0$ are bilepton gauge bosons.

A global symmetry ${\cal L}$ of the full Lagrangean, not broken by
the VEV of the scalars, which commutes with the $SU(3)_L\otimes
U(1)_X$ Gauge group, can be constructed via the
equation~\cite{lon, tully}

\begin{equation}\label{lll}
{\rm L}-\frac{2\lambda_8}{\sqrt{3}}={\cal L}I_3,
\end{equation}
which implies the following assignments: ${\cal
L}(\psi_{lL})=1/3,\; {\cal L}(Q_{iL})=2/3,\; {\cal L}(Q_{3L},\;
\rho)=-2/3,\; {\cal L}(\chi)=4/3,\;{\cal L}(D^c_{iL})=-2,\; {\cal
L}(U_L^c)=2,\; {\cal L}(l^+_L=2$ and ${\cal L}(u_a^c,\;d_a^c)=0$
(the Gauge Bosons are neutral under ${\cal L}$).

Since the scalar sector is very simple now, the model is highly
predictable. As a matter of fact, the full scalar potential
consist only of the following six terms~\cite{ponce}:
\begin{eqnarray}\nonumber
V(\chi,\rho)&=&\mu_1^2|\chi|^2+\mu_2^2|\rho|^2+\kappa_1|\chi^\dagger\chi|^2
+\kappa_2|\rho^\dagger\rho|^2\\
&+&\kappa_3|\chi|^2|\rho|^2+\kappa_4|\chi^\dagger\rho|^2+h.c..
\end{eqnarray}

A simple calculation shows that both, ${\cal L}$ and the lepton
number L are conserved by $V(\chi,\rho)$ and also by the full
Lagrangean, except for the Yukawa interactions ${\cal L}^Y={\cal
L}^Y_{LNC}+{\cal L}^Y_{LNV}$. They induce masses for the fermions
as follows
\begin{eqnarray}\label{yukas}\nonumber
{\cal L}^Y_{LNC}&=&h^U\chi^*Q_{3L}CU_L^c+h^D_{ij}\chi Q_{iL}CD^c_{jL}\\ \nonumber
&+&h^d_a\rho^*Q_{3L}Cd^c_{aL}+h^u_{ia}\rho Q_{iL}Cu_{aL}^c \\
&+&h_{ll^\prime}^e\rho^*\psi_{lL}Cl^{\prime +}_L+h_{ll^\prime}\rho
\psi_{lL}C\psi_{l^\prime L} +h.c.
\end{eqnarray}
\begin{eqnarray}\nonumber
{\cal L}^Y_{LNV}&=& h_a^u\chi^*Q_{3L}Cu_{aL}^c + h^d_{ia}\chi Q_{iL}Cd_{aL}^c \\ 
&+&h_i^D\rho^*Q_{3L}CD_{iL}^c + h^U_{i}\rho Q_{iL}CU_{L}^c +h.c.,
\end{eqnarray}
where the subscripts LNC and LNV indicate lepton number
conserving and lepton number violating term respectively. As a
fact, ${\cal L}^Y_{LNV}$ violates explicitly ${\cal L}$ and L by
two units.

After spontaneous breaking of the gauge symmetry, the scalar
potential $V(\chi,\rho)$ develops the following lepton number
violating terms:
\begin{eqnarray}\label{vlnv}\nonumber
V_{LNV}&=&
v_1[\sqrt{2}H_\chi(\kappa_1|\chi|^2+\kappa_3|\rho|^2)] \\ 
&+&v_1\kappa_4[\rho_ 1^-(\chi^\dagger\rho)
+\rho_1^+(\rho^\dagger\chi)],
\end{eqnarray}
where we have defined as usual
$\chi_1^0=v_1+(H_\chi+iA_\chi)/\sqrt{2}$. $H_\chi$ and $A_\chi$
are the so called CP even and CP odd (scalar and pseudoscalar)
components, and for simplicity we are taking real VEV.

Notice that the lepton number violating part in~(\ref{vlnv}) is
trilinear in the scalar fields, and as expected, $V_{LNV}=0$ for
$v_1=0$. From the former expression we can identify $A_\chi$ as
the only candidate for a Majoron in this model.

The minimization of the scalar potential has been done in full
detail in Refs.~\cite{ponce}. For that purpose two more definitions were
introduced: $\rho_2^0=v+(H_\rho+iA_\rho)/\sqrt{2}$ and
$\chi_3^0=V+(H^\prime_\chi+iA^\prime_\chi)/\sqrt{2}$. An outline
of the main results in Ref.~\cite{ponce}, important for the
present discussion, is:

\begin{itemize}
\item The three CP odd pseudoscalars $A_\chi,\;
A^\prime_\chi$ and $A_\rho$, the would be Goldstone bosons, are
eaten up by $Z,\; Z^\prime$ and $(K^0+\overline{K}^0)/\sqrt{2}$,
the real part of the neutral bilepton gauge boson.
\item Out of the three CP even scalars,
$(v_1H^\prime_\chi-VH_\chi)/\sqrt{v_1^2+V^2}$ becomes a would be
Goldstone boson eaten up by $i(K^0-\overline{K}^0)/\sqrt{2}$, the
imaginary part of the neutral bilepton gauge boson which picks up
L=2 via $H_\chi$. The other two CP even scalars become the SM
Higgs boson and one extra Higgs boson with a heavy mass of order
$V$ respectively.
\item In the charged scalar sector
$(\rho_1^\pm,\; \chi_2^\pm,\; \rho_3^\pm)$ there are four would be
Goldstone bosons, two of them are
$(V\chi_2^\pm-v\rho_3^\pm)/\sqrt{V^2+v^2}$ with L$=\pm 2$, eaten
up by $K^\pm$, and other two with L=0 eaten up by $W^\pm$.
\item Two charged scalars remain as physical states.
\end{itemize}

Counting degrees of freedom tells us that there are in $\chi$ and
$\rho$ twelve ones namely, three neutral CP even, three neutral CP odd and six
charged ones. Eight of them are eaten up by the eight gauge bosons
$W^\pm,\; K^\pm, K^0,\; \overline{K}^0,\; Z$, and $Z^\prime$. Four
scalars remains as physical states, one of them being the SM Higgs
scalar.

Since L is explicitly broken in the context of this model by the
Yukawa term ${\cal L}^Y_{LNV}$, the result is that the would be
pseudo Goldstone Majoron $A_\chi$, the only CP odd electrically
neutral scalar with L=2, is eaten up by
$(K^0+\overline{K}^0)/\sqrt{2}$, the real part of the bilepton
gauge boson.

A variant of this model was introduced in Ref.~\cite{gps} where
the authors considered the fermion mass spectrum under a $Z_2$
discrete symmetry which discards all the LNV interactions in the
Yukawa potential(${\cal L}^Y_{LNV}=0$). For this variant of the model, ${\cal L}$ is
conserved through the entire Lagrangean, the lepton number L is
only spontaneously violated by $V_{LNV}$ in Eq.(~\ref{vlnv}) and the
would be Majoron $A_\chi$ is gauged away, eaten up by
$(K^0+\overline{K}^0)/\sqrt{2}$. Notice that being ${\cal L}$ a
good quantum number now, the spontaneous violation of $SU(3)_L$
implies the spontaneous violation of L via Eq.~(\ref{lll}),
something that it is now allowed because the fermion sector
for L is vectorlike and thus non-anomalous. As attractive as the model is by itself, it is in someway incomplete. In fact, as a consequence of the $Z_2$ discrete symmetry used, new zero VEV scalar fields must be added in order to reproduce a consistent mass spectrum¿\cite{gps}. The 
quark mass spectrum without making use of the $Z_2$ symmetry is analyzed in Ref.~\cite{malong}.

In this comment, we have identified a model in which the Majoron, coming from the spontaneous violation of the lepton number, is gauged away. This unusual mechanism that we have noticed here by the first time, has its origin in the symmetry breaking process and relation (\ref{lll}).

\end{document}